\input{aipcheck}

\documentclass[
    ,final            
  ]
  {aipproc}

\layoutstyle{8x11double}

\begin{document}

\title{Supersymmetry: The Final Countdown}


\author{Hans Peter Nilles}{
  address={Bethe Center for Theoretical Physics and
Physikalisches Institut der Universit\"at Bonn,\\
Nussallee 12, 53115 Bonn, Germany}
}

\begin{abstract}
 
There is hope that the Large Hadron Collider (LHC) at CERN will tell
us about the fate of supersymmetry at the TeVscale. Therefore
we might try to identify our expectations for the discovery of SUSY,
especially in the first years of operation of this machine. In this talk
we shall concentrate on the simplest SUSY scheme: the MSSM with SUSY
broken in a hidden sector mediated by interactions of gravitational
strength (gravity-, modulus and mirage-mediation). Such a situation
might be favoured in a large class of string inspired models.
There is a good chance to identify such simple schemes by knowing the
properties of the gaugino mass spectrum 
such as the gluino/neutralino mass ratios.

\end{abstract}

\maketitle


\section{Introduction}

SUSY08 is the last conference of this series before the Large Hadron
Collider (LHC) gets into operation. It might answer the question whether
TeV-scale supersymmetry is a myth or reality
\cite{reviews}. We should therefore carefully
analyze our expectations (and predictions) for supersymmetry in the (early)
LHC area. As a guideline for the present talk we shall try to concentrate
on the simplest models and schemes, identify clean signals and rely on
theoretical (and personal) prejudices. We shall focus our attention on the
minimal supersymmetric standard model (MSSM) and basic SUSY breakdown- and
mediation-schemes like gravity, gauge and/or anomaly mediation. If we are
lucky, LHC might become a gluino factory that allows us to extract useful
information to disentagle various schemes via an analysis of the gaugino
mass pattern.
\section{The MSSM}
The MSSM is based on the gauge group $SU(3) \times SU(2) \times U(1)$ with
three families of quarks ($Q, \bar{U}, \bar{D}$) and leptons 
($L,\bar{E}$), a pair of Higgs
fields $(H,\bar{H})$ and a superpotential

\begin{equation}
W= Q\bar{D}H + Q\bar{U}\bar{H} + L\bar{E}H + \mu H\bar{H}.
\label{superpotential}
\end{equation}
The symmetries of the model would allow three more terms in the superpotential
\begin{equation}
Q\bar{D}L + L\bar{E}L + \bar{U}\bar{D}\bar{D}
 \label{violating}
\end{equation}
which are problematic because of Baryon- and Lepton- number violation. As
a low energy effective action the model might contain various soft SUSY
breaking terms \cite{Girardello:1981wz}: 
$m_{1/2}$ for gaugino masses, $m_{0}$ for scalar masses as
well as trilinear (A) and bilinear (B) scalar couplings. The soft terms
should be understood as the result of a spontaneous breakdown of
SUSY at some higher energy scale. This simple scheme 
based on the MSSM makes two
predictions and suffers from two  problems. Attempts to construct a model
with spontaneous SUSY-breakdown have to overcome four obstacles.
\subsection{Two Predictions}
The model predicts gauge coupling unification at a scale
$M_{\rm GUT}\cong 10^{16} GeV$. The gauginos slow down the evolution of
the gauge couplings towards higher energies and 
this fact leads to a GUT scale
larger than in the nonsupersymmetric case (avoiding some problems with
proton decay via dimension-6-operators). The weak mixing angle is predicted
to be $\rm{sin}^2\theta_W\approx 0.23$ for one pair of Higgs fields
$(H,\bar{H})$.

The model predicts an upper bound on the mass of the lightest Higgs boson:
at tree level $m_{h}\leq m_Z$. The reason for this bound is the fact that
the quadrilinear Higgs self-coupling is controlled by the gauge coupling.
Radiative corrections (mainly because of the large top-quark Yukawa
coupling) lead to an increase of the bound  to $m_{h}\sim 130GeV$. A
larger mass for the lightest Higgs boson would rule out the MSSM.

\subsection{Two Problems}

Unlike the Standard Model the MSSM suffers from Baryon- and Lepton-number
violation. In the MSSM, the chiral multiplet for the lepton doublet (L)
and the Higgs doublet (H) have the same quantum numbers and allow the
problematic superpotential terms $(QL\bar{D}, LL\bar{E},
\bar{U}\bar{D}\bar{D})$ that among others would lead to rapid proton decay.
The problem needs a (discrete) symmetry that forbids these terms. The
simplest choice is so-called matter- or R-parity $R_{p}=(-1)^{3B+L+2S}$.
Imposition of such a symmetry makes the lightest SUSY particle (LSP)
stable and provides a convincing candidate for dark matter.

The second problem is connected to the mass term $\mu H \bar{H}$ and is known
as the $\mu$-problem. Such a term is allowed by supersymmetry and we have
to ask the question why $\mu$ is small compared to the GUT-scale. One
might forbid $\mu$ by a symmetry but at the end we need a nontrivial $\mu$
of the order of the soft SUSY-breaking terms. This problem is a serious
challenge to many attempts at realistic model building. Mechanisms to
solve the $\mu$ problem might be distinguished by either new terms in 
the superpotential or new terms in the K\"ahler potential.

\subsection{Four Obstacles}

The soft SUSY-breaking terms of the MSSM should originate from a
spontaneous breakdown of supersymmetry. First attempts at model building
in the 1970's and early 1980's turned out to be more difficult than
expected
\cite{Dine:1981za,Dimopoulos:1981au} 
There are some
reasons for this.

\subsubsection{1. Sum Rule}

The supertrace of the (mass)$^2$-matrix vanishes at tree level for
F-term SUSY breakdown \cite{Ferrara:1979wa}: 
this implies that we cannot push up all scalar
masses in a uniform way. If some scales move up compared to the fermion
mass, others have to become light. D-term breaking might give a positive
contribution to ${\rm STr}{\cal M}^2$, but gauge and/or gravitational
anomalies spoil possible solutions. A suggested way out relies on the fact
that the sum rule holds at tree level only. Thus we have to avoid
mass-splittings at tree level (which have the wrong pattern)
and rely on radiative corrections to lead to an acceptable result. The
SUSY-breakdown sector couples only weakly to the
observable sector and therefore has to be "remote" or "hidden".

\subsubsection{2. Gaugino Masses}
Gaugino masses vanish at the renormalizable (tree) level. A possible
mass term $F\chi\chi$ is a dim-5-operator and would require a coupling
constant $1/M$, where M denotes a new mass scale of the model. A
resolution of the problem requires new physics at the mass scale M and/or
radiative corrections. Again we need a new "remote" or "hidden" sector.

\subsubsection{3. R-Axion}

The MSSM with R-parity
has a global (accidental) $U(1)_{R}$-symmetry for the renormalizable terms
in the action. For R charge $R(\theta)= 1$, we choose $R (H,\bar{H})=1$
and $R (Q,\bar{D}, \bar{U}, L, \bar{E})= 1/2$ 
and the allowed terms in the superpotential
have R-charge 2. This R-symmetry forbids gaugino masses and has thus to be
broken. If such a breakdown is spontaneous this would lead to a Goldstone
boson. In the present case $U(1)_{R}$ has a QCD anomaly and we obtain an
axion. Such an axion does not exist.
To avoid the problem we need an explicit breakdown of $U(1)_{R}$.
This again requires a new hidden sector, only weakly coupled to the MSSM.

\subsubsection{4. Vacuum Energy}

The vacuum energy of spontaneously broken global supersymmetry is
strictly positive $V= \Sigma_i F_{i}F^{*}_{i}+\frac{1}{2} D^{2}$. One might
ignore this problem as it is only a problem 
in the presence of gravity. But it is good to
see that in the framework of spontaneously broken supergravity we can have
a vanishing (or small) cosmological constant.

This is all one has to know in order
to construct a successful model of supersymmetry 
breakdown. One has to overcome the four obstacles mentioned and one has to worry 
about the $\mu$-term (and the value of the soft parameter B). Gaugino masses and the 
A-parameter need a breakdown of the (accidental) R-symmetry $U(1)_{R}$. It is obvious 
from the discussion above that SUSY breakdown should happen in a hidden sector, that 
is only weakly coupled to the observable sector (MSSM). One also needs new physical 
phenomena that connect hidden and observable sector and mediate the breakdown of SUSY 
to the MSSM.

\section{SUSY breakdown and mediation}

From obstacle 4 we know that in some way we should 
include gravitational interactions 
to cancel the vacuum energy. It would thus be a natural 
choice to have gravitational 
interactions to connect hidden and observable sector, a scheme called gravity 
mediation. The original suggestion \cite{Nilles:1982ik}
relied on the mechanism of hidden sector gaugino 
condensation 
$< \chi \chi > \sim \Lambda^3$, where $\Lambda$ is 
the renormalization group invariant 
scale of the hidden sector confining gauge group. 
Supersymmetry is broken with an 
F-term: 
$F \sim \Lambda^3/M_{{\rm Planck}}$ leading to a gravitino mass

\begin{equation}
m_{3/2} \sim \frac{\Lambda^3}{M^{2}_{\rm Planck}}.
\end{equation}
Observable sector gaugino masses are generated via the (nonrenormalizable) 4-fermion 
terms of supergravity

\begin{equation}
\frac{1}{M^{2}_{\rm Planck}}\tilde{g} \tilde{g}\chi \chi 
\end{equation}
and we expect soft SUSY breaking terms of order $m_{3/2}$ in the 
observable sector. 
All of the problems and obstacles are taken care of \cite{Nilles:1982ik}: 
soft scalar 
masses are generated 
via radiative corrections,  $m_{1/2}$ is generated via 
nonrenormalizable 
gravitational interactions that break  $U(1)_{R}$ explicitly to 
avoid the axion and 
the vacuum energy is tunable to the desired value. The $\mu$-term
can be generated by higher order gravitational interactions

This scheme was suggested before the full supergravity action
coupled to matter was known.
Subsequently, the
supergravity action with more than one chiral multiplet has been 
worked out in full 
detail \cite{Cremmer:1982wb}. 
Surprisingly, it revealed a modification of the supertrace formula 
that allowed positive 
(mass)$^2$ terms for all scalars. Properties of the models have 
been worked out in 
1982 \cite{gravitymediation}. 
This lead to the first convincing implementation of 
SUSY breakdown in the 
MSSM and this scheme is since known as gravity mediation. Other attempts 
at SUSY breakdown were 
abandoned for almost a decade.

This scheme gained theoretical support in the framework 
of string theory. Especially 
the $E_8 \times E_8$ heterotic string seemed to be particularly 
suited for a hidden 
sector SUSY breakdown via gaugino condensation \cite{DIN}.
This was a 
new variant of gravity 
mediation in which certain moduli fields played the role of 
messengers between hidden 
and observable sector. This scheme of modulus and/or dilaton mediation is 
characterized via soft terms that are of order of magnitude of 
the gravitino mass. 
Details, of course, depend on the nature of the underlying string 
theory. In some 
cases, e.g. the heterotic theory with T-modulus mediation, soft 
terms were supressed 
at tree level. In this case the soft terms were generated through radiative 
corrections \cite{Ibanez:1986xy}
and turned out to be smaller than the gravitino mass.

Gravity (or modulus) mediation is connected to physics at the 
Planck scale. Since 
gravity is a  nonrenormalizable theory with many possible 
higher dimensional operators, one 
might be worried about the control of these operators and 
their contribution to the 
soft terms. Such a control, of course, has to come from a 
meaningful and consistent 
underlying theory such as string theory. Today one can often 
hear and read statements in the literature 
of the type: ``gravity mediation has a flavour problem", 
because of these 
nonrenormalizable operators that might induce flavor changing 
neutral current through  
the soft terms. Such statements are about as meaningful as 
statements like: ``string 
theory has nothing to do with particle physics". The question 
is not whether a given 
scheme can be inconsistent but rather the search for an 
underlying theory that avoids 
a given potential problem in a natural way. In fact, string 
theory offers many 
convincing ways to avoid the flavor problem of gravity mediation, 
e.g. via discrete 
family symmetries and/or universal mediation schemes such as 
dilaton domination in 
the heterotic theory.

In the early 1990's other schemes called ``gauge mediation" 
\cite{gaugemediation}
were proposed 
(and revived) to solve 
the  ``flavour problem" of gravity mediation by construction. 
These schemes required a 
new hidden sector as well as new messengers at a mass scale smaller than 
$M_{\rm GUT}$ or $M_{\rm Planck}$. As a result these schemes had the universal 
property that 
$m_{\rm soft} \gg m_{3/2}$ and thus 
$m_{3/2} \ll$ TeV. The new sectors contain particles with 
standard model couplings 
and their presence  
might spoil gauge coupling unification (as predicted in the 
MSSM). Unification (if 
present) is no longer a prediction but the outcome of careful 
model building. The 
explicit construction of these models has to face the obstacles 
described earlier, 
such as the R-axion, the $\mu$-and the B-problem. It is 
not obvious till 
today how such a scheme can be realized in a simple and elegant 
form. It seems that the most convincing resolution of the 
problems requires higher order gravitational contributions
\cite{Kim:1983dt,Giudice:1988yz}.

\section{Lessons from LEP}
Today we hope that the LHC will clarify the situation with 
supersymmetry at the 
TeV-scale. We should remember, however, that similar expectations 
existed before the 
start of the LEP machine in 1989. The first stage of LEP did 
not see SUSY particles 
but gave the encouraging sign of gauge coupling unification 
with the resurrection of 
(supersymmetric) grand unified theories. Discovery was expected 
at the second stage: 
LEPII. What we got was a support for grand unification and a lower 
bound on the Higgs 
boson mass $m_n > 114$ GeV. Electroweak precision data is consistent
 with the 
standard model, prefers a rather low Higgs mass and gives a good fit 
to the MSSM (in 
contrast to other alternatives of physics beyond the standard model). 
So we have no 
direct sign for SUSY, but some encouragement. Still, a large 
part of the parameter 
space is gone, So, what keeps us going? Certainly, gauge coupling 
unification of the 
MSSM is encouraging, so is the perfect agreement with electroweak 
precision data. The 
model has a convincing dark matter candidate through the 
LSP-neutralino. The value of 
$(g-2)$ of the muon seems to be at odds with the standard model 
and could easily be a 
first indirect sign of SUSY. But we do not know and we need the 
LHC to judge. But 
what should we expect. After more than 20 years of model 
building we have many, many 
models as well as a plethora of mediation schemes. SUSY 
could manifest 
itself  in various different ways and we have to wait for the
LHC to provide the answer.

\section{Theoretical guide lines} 
Still we hope that nature is kind to us and chooses a simple 
and compelling scheme. 
We thus stick to the MSSM with its properties of unified 
coupling constants. 
Theoretical input will come from grand unification and string 
theory. We want to get 
an idea what we can learn from strings for particle physics. 
Recent progress has come 
from 
\begin{itemize}
\item explicit model building towards the MSSM and the 
framework of the heterotic 
braneworld and the concept of local grand unification \cite{Lebedev:2006kn}
\item moduli stabilization and SUSY breakdown 
\cite{Dasgupta:1999ss,Giddings:2001yu,Becker:2003gq,Gurrieri:2004dt}
and its consequences 
for mediation 
schemes.
\end{itemize}
The importance of the mechanism of the fine tuning of the 
vacuum energy for the soft 
SUSY breaking terms has only been appreciated recently 
\cite{Choi:2004sx,Choi:2005ge}
and is closely connected to
a scheme now known as mirage mediation \cite{LoaizaBrito:2005fa}. 
In its simplest 
form it has been found in the 
framework of type IIB string theory in the presence of 
background 3-form fluxes and 
gaugino condensates, completed with an uplifting sector that 
adjusts the 
cosmological constant to the desired value \cite{Kachru:2003aw}. 
In such a set-up, the superpotential contains 
contributions from fluxes and gaugino condensates

\begin{equation}
W = {\rm flux} - \exp (- X)
\end{equation}
where ``flux" is a small quantity in terms of the Planck mass 
and thus $X$ (related to the vev of a modulus) 
is a moderately large number. In 
fact one obtains
\begin{equation} 
X \sim \log (M_{{\rm Planck}}/m_{3/2})
\end{equation} 
providing a ``little hierarchy", resulting from the 
the appearance of the logarithm of 
the Planck-weak scale 
hierarchy. The tree level contribution from modulus 
mediation is therefore suppressed 
by the factor

\begin{equation}
\log (M_{{\rm Planck}}/ m_{3/2})\approx 4 \pi^2
\end{equation}
and radiative contributions from SUSY breakdown in 
the uplifting sector become 
competitive and lead to a mixed mediation scheme. A simple 
scheme of mediation via 
radiative  corrections is known under the name of anomaly 
mediation \cite{anomalymediation}, where the 
individual soft terms are controlled by the $\beta$- and 
$\gamma$- functions of the 
MSSM. As we will explain in a moment, a mixed anomaly-modulus 
mediation scheme leads 
the phenomenon of mirage mediation \cite{Choi:2005ge,mirage2,mirage3}. 

\subsection{Mirage Mediation}

In such a scheme we have mass relations reflecting the little hierarchy
\begin{equation}
m_X \sim \langle X \rangle m_{3/2} \sim \langle X \rangle^2 m_{soft},
\end{equation}
i.e. gravitino and moduli are heavy, and we obtain a 
characteristic pattern of soft 
breaking terms. To see this let us consider the gaugino masses 
\begin{equation}
M_{1/2} = M_{{\rm modulus}} + M_{{\rm anomaly}}
\end{equation}
as a sum of two contributions of comparable size. 
$M_{{\rm anomaly}}$ is non-universal 
below the GUT-scale and proportional to the $\beta$-function, 
i.e. negative for the 
gluino and positive for wino and bino.

With the evolution of gauge coupling constants as displayed in Fig. 1,
\begin{figure}
  \includegraphics[height=.21\textheight]{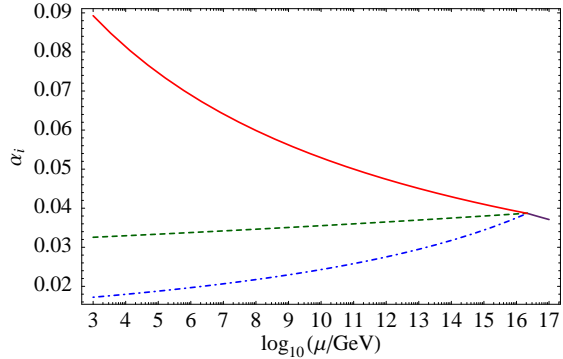}
  \caption{Gauge unification in the MSSM}
\end{figure}
we obtain the evolution of the gaugino masses (Fig. 2)
exhibiting a mirage unification at an intermediate scale.

\begin{figure}
  \includegraphics[height=.21\textheight]{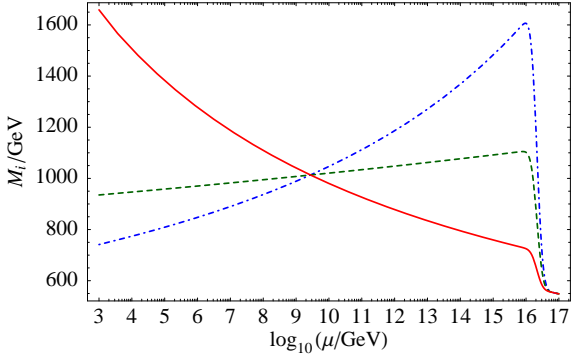}
  \caption{Mirage unification of gaugino masses at an intermediate scale.}
\end{figure}

The splitting of the values at the GUT scale is compensated by the evolution 
(proportional to the same $\beta$-function) and the gaugino 
masses meet at the mirage 
scale
\begin{equation}
\mu_{\rm mirage} \approx M_{\rm GUT} \exp (- 8 \pi^{2}/\varrho),
\end{equation}
where $\varrho$ denotes the ratio of modulus versus anomaly mediation. Thus
\begin{equation}
M_a = M_s (\varrho + b_a g^{2}_{a})= \frac{{m}_{3/2}}{16\pi^2} (\varrho + b_a 
g^{2}_{a}),
\end{equation}
where $\varrho \rightarrow 0$ would correspond to pure anomaly mediation. 
The outcome 
is a very predictive scheme where all the soft terms are determined 
by just two 
parameters: $m_{3/2}$ and $\varrho$. There are constraints on the 
parameters as can be 
seen in Fig. 3.
\begin{figure}
  \includegraphics[height=.22\textheight]{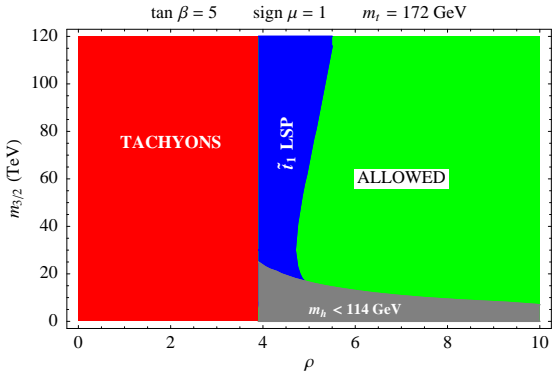}
  \caption{Constraints on $\varrho$ in one of the simplest schemes:
Type IIB string theory with matter on D7 branes. The value of $\varrho$
depends on the uplifting mechanism.}
\end{figure}
For small $\varrho$ one typically has problems with
the presence of tachyons. Even if one removes these tachyons
one has to face problems with either inconsistent candidates
for the (stable) lightest supersymmetric particle (LSP), 
the existence of colour- and
charge-breaking vacua and/or the absence of electroweak symmetry breakdown.
There is also a lower limit on the gravitino mass from the
Higgs mass bound from LEP. But suprisingly, for such a simple
scheme with just two free parameters, a large part of the
parameter space is consistent with all presently know constraints
\cite{mirage4}.

\subsection{The ``MSSM Hierarchy Problem''}

The simplest scheme predicts a rather high scale for the mass of the
gravitino and LSP-neutralino. One might thus be worried about 
the necessity of a (mild)
fine tuning to obtain a rather low mass for the electroweak gauge
bosons less than a hundred GeV,
a fact that has been called the ``MSSM hierarchy problem''
or the ``little hierarchy problem''. Electroweak symmetry breakdown
requires the relation 
\begin{equation}
\frac{m_Z^2}{2}
 =
 -\mu^2+\frac{m_{H}^2-m_{\bar{H}}^2\,\tan^2\beta}{\tan^2\beta-1}
\end{equation}
and there are large corrections to $m_{\bar{H}}^2$.
The influence of the various soft terms is given by
\begin{eqnarray} 
m_Z^2 
 &\simeq& -1.8\,\mu^2 +{ 5.9\, M_3^2 -0.4\, M_2^2 -1.2\, m^2_{H_u}} 
                   +  \nonumber\\
 &&{}+0.7\, m^2_{u_\mathrm{R}^{(3)}}
{ -0.6\, A_t\, M_3} + 0.4\, M_2\, M_3+ \dots\;   
\end{eqnarray}
We see that the gluino mass is the driving force for this evolution.
Mirage mediation improves the situation especially for small
$\varrho$, because of a reduced gluino mass and a ``compressed''
spectrum of the supersymmetric partners. 
Explicit model building (based on the mirage scheme)
towards a solution of this
problem can be found in ref.\cite{finetuning}.

\subsection{Explicit Schemes}

The different schemes differ by the mechanism responsible for the
fine tuning of the vacuum energy, sometimes called the ``uplifting
sector''.

\subsubsection{Uplifting by anti-D3 branes}

This is the original suggestion by KKLT \cite{Kachru:2003aw}
for uplifting 
the vacuum energy in Type IIB string theory with matter on
D7-branes. The scheme leads to $\varrho\sim 5.5$ and a mirage scale
as displayed in Fig. 2. Observe that this value of $\varrho$
is consistent with the constraints of Fig. 3. 
It should be stressed that this scheme
leads to what one might call
pure mirage mediation: gaugino masses and scalar masses
all meet at a common mirage scale.

\subsubsection{Uplifting via matter superpotentials}

Here the uplifting comes from the F-term contribution of a
hidden sector of ``matter'' fields in a hidden sector that
breaks supersymmetry  \cite{LNR}.
It has more free parameters and allows a continuous variation
of $\varrho$. It also leads to potentially new contributions
for the A-parameters and soft scalar mass terms. Gaugino masses
still meet at the mirage scale, but the soft scalar masses
might be dominated by the modulus F-terms and do not show the
same mirage pattern. The constraints on $m_{3/2}$ and $\varrho$
are rather similar to the case of pure mirage mediation as 
displayed in Fig. 4.

\begin{figure}
  \includegraphics[height=.35\textheight]{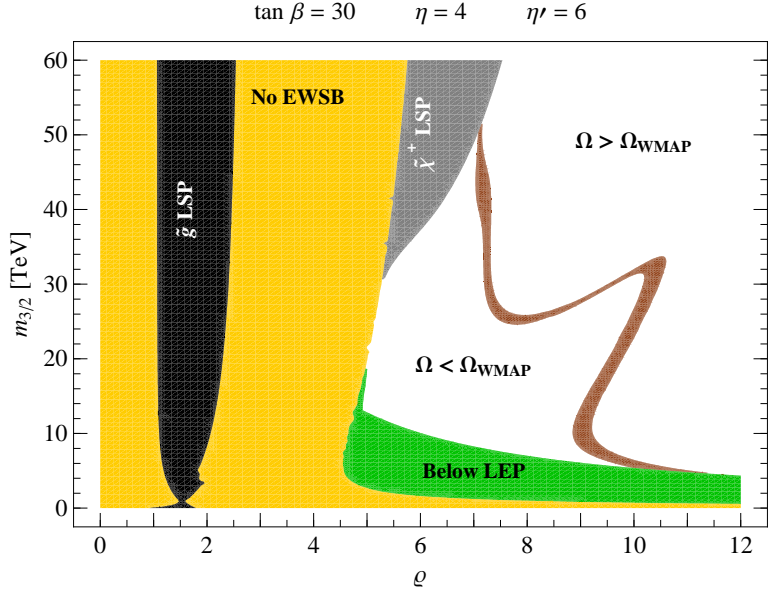}
  \caption{Constraints on $\varrho$ in a scheme of relaxed mirage 
mediation, as found in the heterotic theory with F-term uplifting,
where the dilaton is the relevant modulus for SUSY breakdown
\cite{Lowen:2008fm}.}
\end{figure}

Again there appears a lower limit on $\varrho$ form the
requirement of electroweak symmetry breakdown and the absence
of problematic LSP as well as a lower limit on  $m_{3/2}$ from the
LEP data. The region to the right hand side is allowed and the
dark (brown) strip indicates the region where the abundance of the
LSP-neutralino gives the right amount of dark matter.

\subsubsection{General F-term uplifting}

This ``relaxed'' mirage mediation scheme, where only the
gaugino masses show a mirage pattern is rather common for
F-term uplifting \cite{LLMNR}. 
Still, the ``pure'' mirage pattern is
possible as well, but only for certain values of the parameters.
In general, the patterns for the soft scalar mass terms and
the values of the A-parameters are strongly model dependent.

The main message from here is that gaugino mass predictions
are more robust than predictions for the other soft parameters
and that in the given situation the mirage patters is rather
generic. 

\subsubsection{D-term uplifting}

Uplifting via D-terms alone seems rather problematic. In supergravity
there is the relation
\begin{equation}
D=  \frac{F}{W}
\end{equation}
which vanishes in the KKLT \cite{Kachru:2003aw} minimum with $F=0$ and $W\not= 0$.
Such a minimum can therefore not be uplifted by D-terms alone
\cite{Choi:2005ge}.

In additions we know that generically
\begin{equation}
F\sim m_{3/2} M_{\rm Planck} \ \ \ \  {\rm while} \ \ \ \  D\sim m_{3/2}^2 
\end{equation}
such that for $m_{3/2}\ll M_{\rm Planck}$ the D-terms are
irrelevant \cite{Choi:2006bh}.

\section{The gaugino code}

So, what should we expect to see in the early stage of the LHC
where we scatter protons on protons, i.e. quarks/gluons on 
quarks/gluons. Therefore LHC is a machine designed to produce strongly
interacting particles. If supersymmetry is the physics beyond the
standard model this will lead to a production of squarks and gluinos.
If we are lucky and the mass of the gluino will not be too large,
LHC might become a gluino factory. The produced gluino will decay
in a cascade down to standard model particles and the LSP that might
escape undetected,
but leaving a trace of ``missing energy''. A first step to test
these ideas at the LHC would be the study of the pattern of gaugino
masses \cite{Choi:2007ka}. 
This is particularly interesting as the values for the
gaugino masses show a rather mild model dependence.

To be specific, let us assume models with the particle content
of the MSSM and the measured values of the gauge coupling
constants at the TeV scale. They are approximately given by
\begin{equation}
g_1^2 : g_2^2 : g_3^2 \simeq 1 : 2 : 6.
\end{equation}
The evolution of the couplings would then lead to unification at
a scale around $10^{16}$ GeV (as shown in Fig. 1). 

Observe now that the evolution of the gaugino masses is closely
related to the evolution of the gauge couplings. For the MSSM
the ratio $M_a/g_a^2$ does not run at one loop. This provides rather
robust statements about the gaugino masses and gives us the hope that
gaugino mass relations might be the key to reveal properties of the
underlying scheme of supersymmetry breakdown. We can identify
three basic patterns of gaugino masses.

\subsection{SUGRA Pattern}

In simple schemes like gravity and modulus mediation with universal
soft terms at the GUT scale this pattern would be
\begin{equation}
M_1 : M_2 : M_3 \simeq 1 : 2 : 6 \simeq    g_1^2 : g_2^2 : g_3^2.
\end{equation}
The LSP neutralino $\chi_1^0$ will be 
predominatly a Bino and the mass ratio
\begin{equation}
G=  \frac{M_3}{m_{\chi_1^0}}\simeq 6,
\end{equation}
a characteristic signature for these schemes.

\subsection{Anomaly Pattern}

Just below the GUT scale the gaugino mass ratios are given by
the MSSM $\beta$ functions
\begin{equation}
\beta_1 : \beta_2 : \beta_3 = 33 : 5 : (-15)  
\end{equation}
leading to a mass pattern at the TeV scale
\begin{equation}
M_1 : M_2 : M_3 \simeq 3.3 : 1 : 9 
\end{equation}
as the signal of anomaly mediation.

The LSP neutralino is predominantly Wino and we obtain 
\begin{equation}
G=  \frac{M_3}{m_{\chi_1^0}}\simeq 9,
\end{equation}
a rather high value.

Of course, pure anomaly mediation is problematic because of
tachyonic sleptons. Still, the problems with these tachyons
could be removed by contributions to the soft scalar mass terms
without changing the gaugino mass pattern.
Therefore one should keep this simple pattern in mind.

\subsection{Mirage Pattern}

The mirage pattern seems to be rather generic for the schemes
discussed in the previous section. We have mixed boundary conditions at the
GUT scale and the pattern depends on the parameter 
$\varrho$. For $\varrho\sim 5.5$ we obtain
\begin{equation}
M_1 : M_2 : M_3 \simeq 1 : 1.3 : 2.5 
\end{equation}
while for (the potentially problematic) limiting case
$\varrho\sim 2$ one gets
\begin{equation}
M_1 : M_2 : M_3 \simeq 1 : 1 : 1. 
\end{equation}
The LSP is predominantly bino and we have a ``compressed'' 
spectrum of gaugino masses with a smaller value of
\begin{equation}
G=  \frac{M_3}{m_{\chi_1^0}}< 6.
\end{equation}

With some precision in the determination of the gaugino masses
we might be able to test gaugino mass sum rules. As an example note
that the combination
\begin{equation}
r =  \frac{1}{M_3}(2(M_1+M_2)-M_3) 
\end{equation}
will approximately vanish both in the SUGRA and anomaly pattern.
In this context a nonvanishing $r$ would then be a signal for the 
mirage scheme and allow a determination of $\varrho$.

\subsection{Uncertainties}

These simple schemes have quite some predictive power since
the results are given in terms of MSSM parameters, i.e. 
gauge coupling constants and the $\beta$,$\gamma$-functions. If we are
lucky such a situation is realized in nature and we might learn
a lot just from a study of the gaugino mass patterns.
Of course, there could be uncertainties 
\cite{Kaplunovsky:1994fg} that cannot be
controlled through the parameters of the MSSM. These include
threshold corrections at intermediate scales that are
characteristic for gauge mediation. In string theories we
might have threshold corrections at the GUT scale with
(in principle) unpredictable consequences for the low energy
spectrum. Further uncertainties might arise from contributions
that depend on the K\"ahlerpotential of hidden/observable sector 
fields \cite{Choi:2007ka}.

\subsection{Stringy Expectations}

As we have seen, the mirage pattern has been discovered in the
framework of TypeIIB string theory with matter on D7 branes
\cite{Choi:2005ge}. 
In the same theory for matter on D3 branes one would expect
the anomaly pattern for the gaugino masses. Of course, here
we have to worry about the presence of tachyonic sleptons,
but the spectrum of the scalars is strongly model dependent
and one might remove the tachyons without disturbing the
gaugino mass ratios. In the heterotic theory we face a similar
situation. In the case where the dilaton is the relevant
modulus for SUSY breakdown one would expect a variant of
mirage mediation \cite{Lowen:2008fm}. 
If instead, other moduli than the dilaton
are responsible for SUSY breakdown, the anomaly contribution
might be relevant, probably spoiled by string theory threshold
contributions at the large scale \cite{Ibanez:1986xy}. 
In the recently discussed
M-theory models compactified on manifolds with $G_2$ holonomy
\cite{Acharya:2007rc},
the basic scheme is mirage mediation with potential large
contributions from the K\"ahler-potential terms \cite{Choi:2007ka}
mentioned previously.

\section{Outlook}

It is up to the LHC now to tell us the truth. We might be confident
that the phenomenon of gauge coupling unification will survive the
LHC era. It might also give us the hint how nature solves the
hierarchy problem (if it really cares about it). Supersymmetry
is still one of the favoured schemes for the physics at the 
TeV scale. If it exists we shall have much fun to figure out
the spectrum \cite{Cho:2007fg}
and try to select the underlying models. This might
take some time. But, if we are lucky nature might have chosen a
simple scheme and even at the early stages of the LHC we might
be able to test these ideas. Certainly the pattern of 
the gaugino masses will be a crucial key in this enterprise.


\begin{theacknowledgments}

This work was partially supported by the
European Union 6th framework program MRTN-CT-2004-503069
"Quest for unification", MRTN-CT-2004-005104 "ForcesUniverse",
MRTN-CT-2006-035863 "UniverseNet" and 
SFB-Transregio 33 "The Dark Universe" by Deutsche
Forschungsgemeinschaft (DFG). I would like to thank the
organizers of the SUSY08 conference for support and 
hospitality.

\end{theacknowledgments}




\end{document}